\documentclass[longauthor]{aa}

\usepackage{graphicx}
\usepackage{txfonts}
\begin{document}

   \title{Can circumstellar interaction explain the strange light curve features of Type Ib/c supernovae?}

     \author{Andrea P. Nagy\inst{1,2}
           \and
           Zs{\'o}fia R. Bodola \inst{1}
          }

   \institute{Department of Experimental Physics, Institute of Physics, University of Szeged, D{\'o}m t{\'e}r 9, 6720 Szeged, Hungary
    \and
     ELKH-SZTE Stellar Astrophysics Research Group, Szegedi {\'u}t, Kt. 766, 6500 Baja, Hungary \\
       \email{nagyandi@titan.physx.u-szeged.hu}\\ }

   \date{Accepted January 24, 2025}


  \abstract
   {The evolution and the surrounding of stripped-envelope supernova progenitors are still under debate: some studies suggest single-star, while others prefer massive binary progenitors. Moreover, the basic physical properties of the exploding star and its interaction with circumstellar matter could significantly modify the overall light curve features of these objects. To better understand the effect of stellar evolution and circumstellar interaction, systematic hydrodynamic calculations are needed.}
   {Here, we test the hypothesis that circumstellar matter generated by an extreme episodic $\eta$ Carinae-like eruption that occurs days or weeks before the supernova explosion may explain the controversies related to the general light curve features of stripped-envelope supernovae. }
   {In this work, we present our bolometric light curve calculations of both single- and binary progenitors generated by hydrodynamic simulations via \textbf{MESA} and \textbf{SNEC}. We also studied the effect of an interaction with a close, low-mass circumstellar matter assumed to be created just a few days or weeks before the explosion. Beyond generating a model light curve grid, we compared our results with some observational data.}
   {We found that merely the shape of the supernova light curve could indicate that the cataclysmic death of the massive star happened in a binary system or was related to the explosion of a single star. Moreover, our study also shows that a confined dense circumstellar matter may be responsible for the strange light curve features (bumps, re-brightening, or steeper tail) of some Type Ib/c supernovae.}
   {}


   \titlerunning{Can CSM interaction explain the strange light curve features of Type Ib/c SNe?}
   \maketitle

%

\section{Introduction}
Core-collapse supernovae (SNe) are formed from the gravitational collapse of the nickel-iron core of massive stars. Despite the same explosion mechanism, these transients could be extremely different from an observational point of view: we can distinguish H-rich (Type IIP or IIL), H-poor (Type IIb), and H-free (Type Ib/c) explosions.
The group of Type IIb together with Type Ib/c is also called stripped-envelope supernovae (SESNe). These noticeable differences in spectral and light curve (LC) features mainly depend on the mass-loss history of the progenitor star \citep[e.g.][]{Limongi, Vink}.

The light curves of SESNe are mainly powered by the radioactive decay of nickel and cobalt as well as gamma-ray trapping \citep[e.g.][]{Arnett, Clocchiatti}. The estimated nickel mass of these explosions is around 0.03–0.35 $M_\odot$ \citep{Taddia1}. According to both analytic and hydrodynamic calculations, their progenitors are quite compact ($R_p \sim 10^9 - 10^{10}$ cm), and the early luminosity of these events are mainly dependent on the progenitor radius. The ejected masses during the explosion are relatively low ($M_{ej} \sim 1.1 - 10.1\ M_\odot$), while the explosion energy of these object are about $0.25 - 4.9 \cdot 10^{51}$ erg \citep[e.g.][]{Lyman, Taddia1}. In some cases an initial peak can be detected, followed by a rapid drop in luminosity. This early emission is believed to be driven by the shock-breakout or an extended envelope around the progenitor \citep[e.g.][]{Bersten1, Nagy1}.

The explosion may occur within a circumstellar matter (CSM) formed throughout stellar evolution, even though some supernova subtypes (IIn, Ibn, Icn) show significant CSM interaction features both in their spectra and luminosities \citep{Fraser, Dessart22, Maeda22, Takei}. The circumstellar matter could play an important role in generating the light curve of some superluminous SNe, even if there are no obvious spectral signs of the interaction \citep{Moriya2, Mazzali, Wheeler2}. Moreover, as \citet{Kuncarayakti} revealed, normal Type Ic supernovae may show similar spectroscopic features to interacting Type Ibn and Icn SNe at late phases. These phenomena, and the fact that we do not know exactly when massive stars, especially stripped-envelope supernova (SESN) progenitors, get rid of their outer hydrogen and/or helium layers, suggest a possible circumstellar interaction for Type Ib/c supernovae as well.

As far as we know, SESN progenitors go through significant mass-loss during the pre-supernova evolution, but the exact mechanism is still under debate. Some observations suggest \citep[e.g.][]{Cao} a massive single star (possibly Wolf-Rayet) progenitor that loses its outer envelope due to extreme stellar wind or irregular eruption phases. However, other studies \citep[e.g.][]{Sana, Woosley} assume binary interaction before the explosion that strips away the outermost layers of the massive donor star. The commonality in both scenarios is that they suggest circumstellar matter around the progenitor star.

Even though recent studies find a possible connection between CSM interaction and the re-brightening of late-time (200$\lesssim$ days after the explosion) SESNe light curves \citep{Sollerman, Soraisam, Kuncarayakti1}, these results indicate a circumstellar ring with a minimal inner edge around  $5\cdot10^{15} - 10^{16}$ cm. On the other hand, theoretical considerations \citep[][e.g.]{Benetti, Maeda21} suggest the CSM radius around a stripped-envelope supernova progenitor should be $10^{14} - 10^{15}$ cm, which presumes that this matter is just ejected a few months before the SN explosion. This controversy may be resolved if we assume extreme episodic mass-loss (0.1 - 1 $M_\odot$) event at the end of stellar evolution may caused by somewhat similar physical processes as we can observe in the case of $\eta$ Carinae \citep{Vamcatira}. This theory may also explain the earlier (around 60-100 days after the explosion) light curve bumps of SESNe if an eruption occurs days or weeks before the supernova explosion. In fact, such an event is more than plausible as precursor outbursts shortly before the explosion have been observed in some cases. For example, an optical transient in 2004 can be connected to the Type Ic explosion SN 2006jc \citep{Pastorello}, while according to \cite{Ho} the progenitor of SN 2018gep also produced outbursts just days to months before the explosion. Moreover, observational data also suggest that precursor outbursts could be common  but less energetic and short-lasting for types other than IIn \citep{Strotjohann}.

In binary systems, besides such an intense eruption \citep{Mcley}, if both the primary and the secondary components have strong stellar wind, a colliding wind structure (CWS) \citep[e.g.][ and reference within]{Kashi} can be formed. This dense formation may mimic a close CSM around the star as supernova ejecta interacts with it, as suggested by \citet{Kochanek} for the H-poor (Type IIb) stripped-envelope supernova, iPTF13as (SN 2013cu).

In these cases, the extra luminosity excess is due to an additional power source related to the interaction between the SN ejecta and a close CSM \citep[e.g.][]{Chevalier}, as the mass-loss history shortly before the SN explosion can drastically influence the optical light curve properties (e.g. brightness and color) of Type Ib/c progenitors \citep{Jung}.

Accordingly, numerical studies could be crucial to reveal the nature of this early CSM interaction and expose a possible mass-loss episode shortly before the supernova explosion. For this, we started our work by adding an attached, relatively thin CSM layer much closer than in the previous works aiming to model type Ibn/Icn explosions \citep[e.g.][]{Kuriyama, Maeda22, Takei}. The radius of our CSM models was defined by multiplying the radius of the progenitor models with factors from 2 to 10. This way, since our model progenitor stars are compact, the radius of the attached CSM does not exceed 10 $R_{\odot}$ in any case. We apply one-dimensional hydrodynamic calculations to generate the bolometric light curves from SESN progenitor models interacting with a close, low-mass ($M_{CSM} \leq 2 M_\odot$) circumstellar matter. In this paper, we introduce our systematic studies related to the effect of the radius and the mass of the CSM, as well as the different physical compositions of the progenitors.

This paper is organized as follows: in Section 2. we present our model setups and numerical method. In Section 3. we discuss the most important properties of our synthetic light curves. We compare our result with some well-known SESNe in Section 4. Finally, in Section 5. we provide a summary of our conclusion.

\section{Model Setups}

We perform complete hydrodynamical modeling and analytic approximations to create the unique physical configuration, self-consistent with a supernova explosion that occurs in a close circumstellar matter caused by an extreme mass-loss event just a few days or weeks before the cataclysm.

We calculate both single star and binary progenitor models using the Modules for Experiments in Stellar Astrophysics (\textbf{MESA} version r-12778), which is a 1-dimensional, numerical hydrodynamic stellar evolution code \citep{Paxton1, Paxton2, Paxton3, Paxton4, Paxton5, Jermyn}. Then, we generate different thin ($R_{CSM} \leq 10\, R_p$), low-mass ($M_{CSM} \leq 2 M_\odot$) CSM configurations with a power-law density profile analytically, and add them to the \textbf{MESA} models. As a final step, we calculate the bolometric light curves of our progenitor models with and without the attached circumstellar matter using the 1D spherical Lagrangian SuperNova Explosion Code \citep[\textbf{SNEC},][]{Morozova}.

\subsection{Progenitor Models}
The detailed explanation of the progenitor models is beyond the scope of this paper (for more specifications see our forthcoming paper). Here, we just show the basic simulation setup to numerically generate a proper environment for our assumed CSM-interacting SESNe to test the effect of a close, low-mass circumstellar matter.

\subsubsection{Single star models}
First, we create stellar models for single massive stars to create Type Ibc SN progenitors. Our calculations run from the zero-age main sequence (ZAMS) to the end of helium burning. Then, we adopt a manually adjusted \textit{mass\_change} parameter with a value of $10^{-2}$ M$_\odot$/yr to remove the remaining hydrogen (and helium) layers before Fe-core collapse. This average mass-loss rate is around the same order-of-magnitude as calculated for the great 1840 eruption of $\eta$ Carinae \citep{Andriesse}. Thus, from a modeling point of view, this more extreme mass-loss phase could correspond to an intensive late-time outburst, which is self-consistent with our CSM-forming premise.

We take stellar evolution models with the initial masses of 15, 20, and 50 M$_\odot$, corresponding within the estimated mass range of SESN progenitors \citep{Georgy}. In all 14 models, we assume semi-convection (\textit{alpha\_semiconvection} = 0.01), overshooting with $f_0 = 0.0005$, and mixing (\textit{mixing\_length\_alpha} = 1.5), while set the \textit{varcontrol\_target} = $10^{-3}$ for the convergence of models with higher mass- loss. To avoid numerical problems and speed up our calculations, we change the nuclear reaction rate network from 'basic.net' to 'approx21.net' with Heger-style adaptive network option \citep{Woosley1}. Besides, we use the \textit{Ledoux} convection criterion to determine the position of the non-radiative boundaries within our stellar model.

\begin{table*}
\centering
\caption{Physical properties of our single star models}
\label{tab:single_models}
\begin{tabular}{lcccccccc}
  \hline
  \noalign{\smallskip}
  No. & $M$ ($M_{\odot}$) & Wind type & $\eta_{wind}$ &  Z & $R_p$ ($10^{10}$ cm) & $M_{Ibc}$ ($M_{\odot}$) & $dM$ ($M_{\odot}$) & $M_{He}$ ($M_{\odot}$)\\
 \noalign{\smallskip}
  \hline
  S1 & 15 & Dutch\tablefootmark{1} & 0.8 & 0.02 & 8.93 & 4.28 & 10.72 & 1.81\\
  S2 & 15 & Dutch\tablefootmark{1} & 0.8 & 0.002 & 4.39 & 4.34  & 10.66 & 1.77 \\
  S3 & 15 & Dutch\tablefootmark{1} & 1.0 & 0.02 & 8.45 & 4.36& 10.64 & 1.83\\
  S4 & 15 & Reimers\tablefootmark{2} & 0.8 & 0.02 & 8.71 & 4.23 & 10.77 & 1.84\\
  S5 & 20 & Dutch\tablefootmark{1} & 0.8 & 0.02 & 7.13 & 6.65  & 13.35 & 3.32\\
  S6 & 20 & Dutch\tablefootmark{1} & 0.8 & 0.002 & 5.52 & 6.73 & 13.27 & 3.72\\
  S7 & 20 & Dutch\tablefootmark{1} & 1.0 & 0.02 & 7.15 & 6.80 & 13.20 & 3.46\\
  S8 & 20 & Dutch\tablefootmark{1} & 0.6 & 0.02 & 8.69 & 6.77 & 13.23 & 3.42\\
  S9 & 20 & Reimers\tablefootmark{2} & 0.8 & 0.02 & 7.17 & 6.82 & 13.18 & 3.29 \\
  S10 & 20 & de Jager\tablefootmark{3} & 0.8 & 0.02 & 7.13 & 6.74 & 13.26 & 3.82 \\
  S11 & 20 & Vink\tablefootmark{4} & 0.8 & 0.02 &  9.89 & 6.35 &  13.65 & 3.39 \\
  S12 & 50 & Dutch\tablefootmark{1} & 0.8 & 0.02 & 0.22 & 6.27 & 43.73 & 0.078\\
  S13 & 50 & Dutch\tablefootmark{1} & 0.8 & 0.002 & 0.24 & 6.11 & 43.89 & 0.098 \\
  S14 & 50 & de Jager\tablefootmark{3} & 0.8 & 0.02 & 0.03 & 6.23 & 43.77 & 0.088 \\
\hline
\end{tabular}
\tablebib{(1)~\citet{Glebbeek, Nugis}; (2)~\citet{Reimers}, (3)~\citet{deJager}, (4)~\citet{Vink1}.}
\end{table*}

In our model grid, we systematically studied some modeling parameters (related to metallicity, stellar wind type, and its scaling factor) that can alter the physical properties and mass-loss of the progenitor. For each mass, the reference parameters are $Z = 0.02$, 'Dutch' stellar wind schemes \citep{Glebbeek, Nugis} for the Asymptotic Giant Branch phase, and $\eta_{dutch} = 0.8$ for the wind scaling factor. Then, we one-by-one vary these parameters to examine their effect. Some properties of the progenitors are provided in Table~\ref{tab:single_models}, giving details of the initial ($M$) and final masses and radius of the progenitors ($M_{Ibc}$ and $R_p$) and its He-mass ($M_{He}$) after the assumed intensive mass-loss phase that creates Type Ibc explosions, as well as the total mass-loss (dM) caused by stellar wind and the \textit{mass\_change} parameter.

\subsubsection{Binary star models}
Here, we mainly focus on close binary systems (initial rotational period, $P_{init}$ $ < 100$ days), where we consider both the primary and secondary components as massive stars. We also assume that higher initial mass ($M_{init}$) relates to the donor star, which evolves to a SESN progenitor due to mass-loss. In our total of 16 models, we vary the initial mass of the donor within 35 - 60 $M_{\odot}$ while altering $M_{init}$ for the secondary component as 12 - 42 $M_{\odot}$. We also use different metallicities (Z), initial rotational periods, and mass ratios (q) to describe the binary system. We also determine the total mass-loss for the donor (dM) and the mass of its remaining He-layer ($M_{He}$) at the time of the core-collapse (Table~\ref{tab:binary_models}.). All models show some amount of He, but considering the synthetic spectral studies of \cite{Hachinger}, some of them may form He-free spectra as approximately 0.06 - 0.14 $M_\odot$ of helium can be hidden and lead to Type Ic classification. Thus, we can classify our binary models as possible Type Ib/c progenitors with a final mass of $M_{Ibc}$ also listed in Table~\ref{tab:binary_models}.

\begin{table*}
\centering
\caption{Physical properties of our binary star models}
\label{tab:binary_models}
\begin{tabular}{lccccccccc}
  \hline
  \noalign{\smallskip}
  No. & $M_{init,1}$ ($M_{\odot}$) & $M_{init,2}$ ($M_{\odot}$) & q &  $P_{init}$ (days) & Z & $R_p$ ($10^{10}$ cm) & $M_{Ibc}$ ($M_{\odot}$) & $dM$ ($M_{\odot}$) & $M_{He}$ ($M_{\odot}$) \\
 \noalign{\smallskip}
  \hline
  B1 & 35 & 12 & 0.343 & 30 & 0.004 & 5.18 & 8.04 & 26.96 & 1.0 \\
  B2 & 35 & 30 & 0.857 & 4 & 0.014 & 3.05 & 3.23 & 31.77 & 0.14 \\
  B3 & 35 & 30 & 0.857 & 50 & 0.014 & 2.83 & 3.18 & 31.82 & 0.14 \\
  B4 & 40 & 15 & 0.375 & 15 & 0.004 & 3.33 & 7.73 & 32.27 & 0.41 \\
  B5 & 42 & 13 & 0.310 & 5 & 0.004 & 2.82 & 7.77 & 34.23 & 0.21 \\
  B6 & 42 & 13 & 0.310 & 15 & 0.004 & 2.39 & 6.66 & 35.34 & 0.061 \\
  B7 & 42 & 13 & 0.310 & 15 & 0.005 & 3.65 & 5.09 & 36.91 & 0.13 \\
  B8 & 42 & 13 & 0.310 & 16 & 0.004 & 2.56 & 6.68 & 35.32 & 0.069 \\
  B9 & 43 & 13 & 0.302 & 10 & 0.004 & 2.58 & 6.46 & 36.54 & 0.059 \\
  B10 & 45 & 13 & 0.289 & 15 & 0.004 & 2.82 & 7.45 & 37.55 & 0.12 \\
  B11 & 60 & 13 & 0.217 & 5 & 0.004 & 4.13 & 5.38 & 54.62 & 0.23 \\
  B12 & 60 & 42 & 0.700 & 5 & 0.007 & 4.22 & 4.58 & 55.42 & 0.26 \\
  B13 & 60 & 42 & 0.700 & 5 & 0.005 & 4.09 & 4.87 & 55.13 & 0.25 \\
  B14 & 60 & 42 & 0.700 & 10 & 0.004 & 4.05 & 5.36 & 54.64 & 0.24 \\
  B15 & 60 & 42 & 0.700 & 15 & 0.004 & 3.998 & 5.33 & 54.67 & 0.23 \\
  B16 & 60 & 42 & 0.700 & 20 & 0.004 & 3.79 & 5.41 & 54.59 & 0.26 \\
\hline
\end{tabular}
\end{table*}

We also select the parameter space of our binary models so that they can be initialized as circular systems by \textbf{MESA}. As in this case, the code can monitor the changes in the orbital angular momentum ($J_{orb}$) during binary evolution. It is essential, as heavy mass transfer, among others (gravitational waves, magnetic braking, and spin-orbit coupling), strongly affects the orbit of the binary system. So, with this approximation, we can create self-consistent binary models that take the alterations of $J_{orb}$ into account \citep[see Eq. 3. in][]{Paxton3}.

We considered a fully non-conservative mass transfer for our binary models, that can be represented in the \textbf{MESA} with a zero \textit{mass\_transfer\_gamma} and \textit{mass\_transfer\_delta} value. At the same time, we fixed \textit{mass\_transfer\_alpha} and  \textit{mass\_transfer\_beta} parameters as 0.5 \citep{Petrovic, Shao}.

All simulations start from the zero-age main sequence and follow the evolution of the binary system till the donor evolves near the core-collapse phase. Nevertheless, none of them reached this state due to numerical reasons. However, most of the Ne is created in the center (except model 9. - 10.) or even forms a Si-core, which is only years to days before the formation of the iron core. Besides reaching core-collapse in a binary, we aim to generate potential progenitor models for stripped-envelope supernovae. Thus, our models need to lose their entire H- and even a good portion of their He-layers, and their final mass could be in good agreement with previous studies suggesting ejecta with 2 - 10 $M_\odot$ \citep[e.g.][ and reference within]{Jin1}. To do so, alongside Roche-lobe overflow, we initialize our \textbf{MESA} models with stellar winds using the "Dutch" hot wind-scheme option with scaling factor $\eta_{Dutch} = 1.0$ and switch on the super-Eddington wind at later evolutionary phases. When some extra energy (via unstable fusion, wave heating, or a binary companion) heats the near-surface region of the star, the stellar photosphere may exceed the Eddington luminosity, resulting in mass-loss \citep{Quataert} called super-Eddington wind that can surpass the effect of the metallicity-driven stellar wind. Note that from a modeling point of view, using super-Eddington wind setups in \textbf{MESA} is capable of controlling additional mass-loss, such as \textit{mass\_change} parameter do for the single star models. However, super-Eddington wind configuration describes a realistic physical phenomenon, while \textit{mass\_change} parameter indicates a fixed, arbitrary mass-loss.
So, to create a self-consistent physical configuration for our binary models, we implement the super-Eddington wind scheme into our simulations.  Here, we also implemented the default values for both cont\-rol parameters: \textit{super\_eddington\_wind\_Ledd\_factor = 1} and \textit{super\_eddington\_scaling\_factor = 1}. In this manner, we followed the classic theoretical condition. Namely, mass-loss starts when the effective surface luminosity surpasses the Eddington luminosity.

 However, as e.g. \citet[]{Pauli} showed, the mass transfer to a close companion star should be the real engine of the heavy mass-loss. Previous studies also suggest that Wolf-Rayet stars in binary systems tend to occur in low-metallicity environments. Thus, the effect of the metallicity-driven wind should be negligible compared to the mass transfer between the two components. Considering these statements, we mainly focus on low-metallicity models in our study.

In \textbf{MESA}, two different approaches are available for modeling the acceptor star. We can assume that a secondary component remains a steady mass point. Or this object can evolve in time like the donor star. Here, we tested both scenarios, and no significant difference is noticeable in the physical properties of the primary component. The main reason for this could be that regardless of $M_{init,2}$, the direction of the mass transfer does not change during the simulation, and the acceptor remains in the same evolutionary stage. Thus, the secondary component has no direct impact on the inner structure of the donor star.

\subsection{CSM Properties}
\begin{figure}[]
\centering
\includegraphics[width=0.48\textwidth]{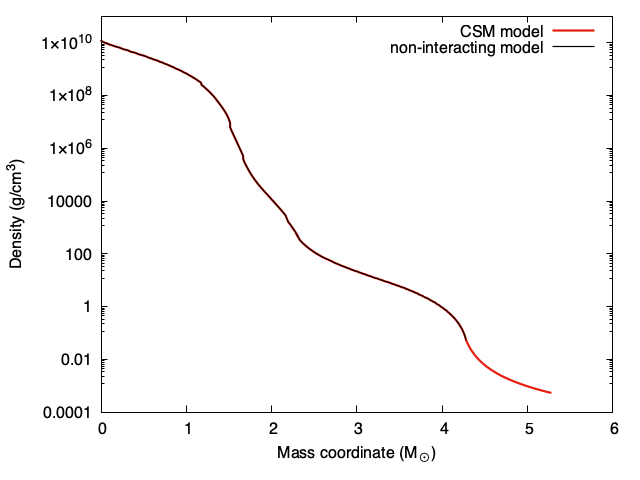}
\caption{Comparison of the initial density profile of a non-interacting and an interacting single-star model. The black line represents the non-interacting reference model, while the red line shows the total density profile of the same model with an added CSM profile.}
\label{fig:dens}
\end{figure}

In the literature, some complex density profiles are available to model the structure of the circumstellar matter. One possibility is a double power-law density profile by \cite{Tsuna}, which describes the transition between the bound and unbound parts of a close CSM attached to the progenitor star. However, this scenario only explains the CSM generated by the eruptions of a single star. Thus, making sure that the single-star and binary models could be comparable, we use a simplified CSM structure to examine the effect of a close, dense material cloud around the exploding stars. We assume that the circumstellar matter is attached to the progenitor. Thus, the inner radius of it is equal to the radius of the progenitor. We adopt a power-law density profile, where the initial density of the CSM ($\rho_0$) is estimated to be identical to the density at the progenitor surface (Fig. \ref{fig:dens}) and its value proportionate to the quadratic of the radius element of the CSM (r) as
\begin{equation}
    \rho(r) = \rho_0\, \left(\frac{R_p}{r}\right)^2\ .
\end{equation}

Here, we also assume that the mass of the CSM is independent of the CSM radius, as these both are grid parameters of the numerical simulation \citep{Tsuna}. Indicating that the $M_{CSM}$ parameter can not be derived from the average mass-loss rate and wind velocity \citep{Chevalier1}, but it is just an arbitrary model parameter. Despite all that, we estimate a 2000 $km/s$ average wind velocity as the velocity of our CSM configuration.

\begin{figure}[]
\centering
\includegraphics[width=0.48\textwidth]{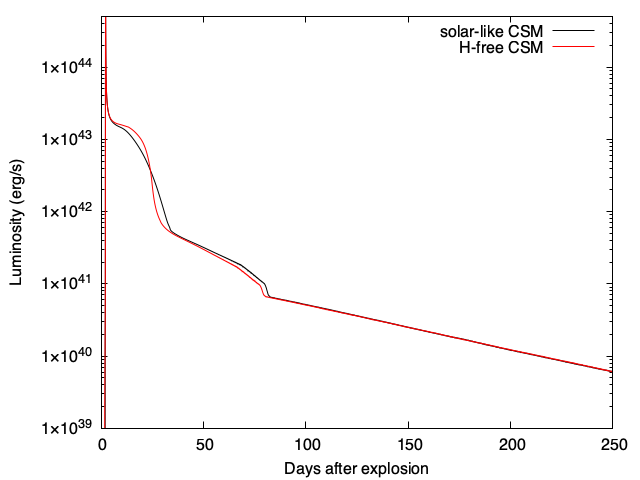}
\caption{Comparison of the calculated bolometric light curves for interacting CSM models ($M_{CSM} = 2 M_\odot$,$R_{CSM} = 10 R_p$ )with different abundances. The red curve represents the H-free CSM, while the black line is related to a solar-like chemical composition.}
\label{fig:abund}
\end{figure}

The chemical composition of the circumstellar matter is assumed to be solar-like, mainly containing hydrogen and helium. A pure He-composition would be more realistic as an expected blown-off layer of a massive, convective star in such a late evolution phase. However, \textbf{SNEC} calculations become extremely time-consuming, and in many cases, numerically unstable for an H-free CSM. Moreover, no significant differences can detected for the overall light curve features or maximal luminosities, except the first peak shows a steeper luminosity cut (Fig. \ref{fig:abund}). Thus, as we are only interested in the general LC characteristics, we chose to apply a solar-like chemical composition for our CSM structures.

The radius of the confined CSM is constrained to be less than 10 times the progenitor radius, while $M_{CSM}$ is at most 2 $M_\odot$. In our model grid, we adopt $R_{CSM}$ of 2, 5, and 10 $R_p$, and CSM mass of 0.01, 0.05, 0.1, 0.5, 1, and 2 $M_\odot$ as an addition for all progenitor models. For single-star models, these assumptions correspond well with the last-stage mass-loss history (controlled by \textit{mass\_change} parameter) of the progenitor stars. On the other hand, we have no modeling evidence for the mass-loss history of our binary configurations, as none of them reached the core-collapse phase. So, from a theoretical point of view, it seems reasonable to assume a similar CSM structure for binaries as used for single-star models.

\subsection{Explosion Properties}
Being a radiation hydrodynamic code, \textbf{SNEC} \citep{Morozova} can take the interaction between SN ejecta and CSM into account in computing bolometric light curves, which is needed to compare our results with observational data, whose bolometric LCs contain unexpected luminosity variations (bumps or re-brightening). To test the effect of a CSM interaction, we simulate the explosion of all the previously described progenitor models with and without a mounted CSM shell.

\begin{figure}[]
\centering
\includegraphics[width=0.48\textwidth]{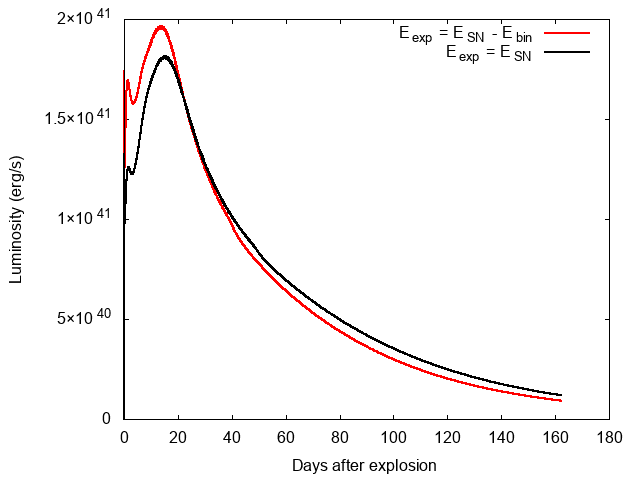}
\caption{Comparison of calculated bolometric light curves with different 'Bomb\_mode' options. The red curve represents 'Bomb\_mode = 1', where $E_{exp} = E_{SN} - E_{bin}$, while the black lines show the fix explosion energy option ('Bomb\_mode = 2').}
\label{fig:bomb}
\end{figure}

We adopt fixed  excited mass ($M_{ex} = 1.4 M_{\odot}$  suggested by \cite{Morozova}), explosion energy ($E_{exp} = 1.5\ \cdot 10^{51}$ erg) and nickel masses ($M_{Ni} = 0.1 M_{\odot}$) to reduce the strong impact of these parameters that can alter our results. Unfortunately, the effect of nickel can not be eliminated.  Because \textbf{SNEC} sums up this implemented nickel mass value with the nickel content of the initial interacting or non-interacting stellar model, which is approximately $10^{-8} - 10^{-9} M_\odot$ and $0.025 - 0.03 M_\odot$ for binary and single-star models, respectively. Moreover, not just the nickel content but the distribution of $^{56}$Ni and the overall density profile of our stellar models can also affect the gained light curves. Thus, if we fix the $M_{Ni}$ parameter, the tail of the light curve still depends on the initial nickel mass that is generated via stellar evolution. To do so, we adopt a $M_{Ni}$ value that corresponds well with both the observations and the theoretical models. Previous observations suggest that the typical nickel mass for SESNe should be about $0.05-0.5 M_{\odot}$ \citep{Lyman, Taddia}, while recent studies indicate a range of $0.06-0.132 M_{\odot}$ \citep{Rodriguez}. While from a theoretical point of view, the expected nickel mass for these objects should be around $0.01-0.1 M_{\odot}$ \citep{Muller, Ouchi, Dessart22}. So, our selected $M_{Ni}$ value could be reasonable for both observational and theoretical aspects. The chosen $E_{exp}$ value is within the typical explosion energy range of  Type Ib/c supernovae gained from both analytic and hydrodynamic calculations. To define the same explosion energy, we should set the 'Bomb\_mode' option accordingly. As a default, \textbf{SNEC} determines the explosion energy as the total energy of the supernova environment. Thus, the code subtracts the binding energy of the exploding star ($E_{bin} < 0$) from the inserted thermal bomb energy ($E_{SN}$). So, the actual explosion energy will be $E_{exp} = E_{SN} - E_{bin}$ that can differ from model to model as it corresponds to the asymptotic energy of the system (SNEC users' manual\footnote{\url{https://stellarcollapse.org/codes/snec_notes-1.00.pdf}}). However, if we use the non-default 'Bomb\_mode = 2' option, the explosion energy is assumed to be equal to the specified thermal bomb energy parameter ($E_{exp} = E_{SN}$). For better understanding, the difference between the gained bolometric light curves using the two optional 'Bomb\_mode' options can be seen in Figure \ref{fig:bomb}. All-in-all, for more self-consistent comparison, the non-default ('Bomb\_mode = 2') setting could be a reasonable choice as it allows to fix the explosion energies. Thus, in further calculations, we used this option.

.

Note that the solver controls, opacity option, and equation of state (EOS) were set as defaults during our calculations. \textbf{SNEC} calculates the opacity of each grid point from Rosseland mean opacity tables that consider the chemical composition, temperature, and density of matter. In addition, it requires the so-called opacity floor parameters that by default 0.01 and 0.24 for the envelope ($Z < 1$) and the core (Z = 1), respectively. These specific opacity floor values are based on the calculations by \cite{Bersten} providing a calibration between LTE and multi-frequency codes. On the other hand, in \textbf{SNEC} the equation of state calculations adopt simplified analytic functions. By default these equations are based on the study of \cite{Paczynski}, which determine EOS for a mixture of ions, photons, and semi-relativistic electrons.

\section{Bolometric Light Curves}
\begin{figure}[!h]
\centering
\includegraphics[width=0.48\textwidth]{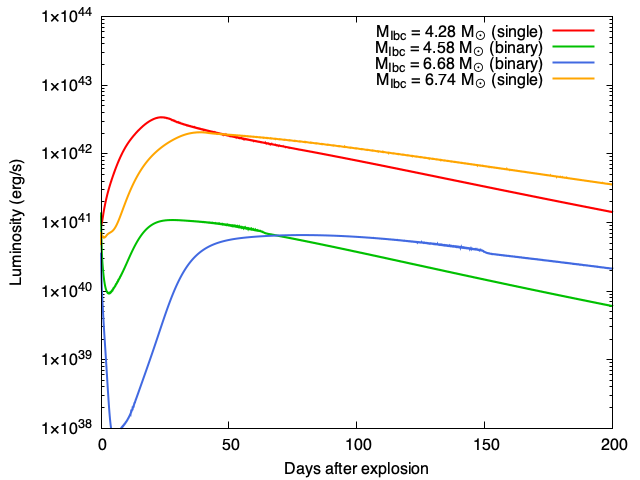}
\caption{Comparison of single-star and binary models having similar progenitor mass. The red and orange lines represent single-star models with 4.28 and 6.74 $M_\odot$, while green and blue ones perform binary calculations with 4.58 and 6.68 $M_\odot$, respectively.}
\label{fig:comp}
\end{figure}

The bolometric light curves of all non-interacting progenitor models show some general features of Type Ib/c supernova explosions. However, the ones generated from binary models typically produce fainter light curves with broader peaks than single star-related ones with around the same progenitor mass, explosion energies, and nickel masses (see Fig. \ref{fig:comp}). This phenomenon may occur due to the different compactness and initial nickel masses of single- and binary progenitors with similar masses. Comparing the $R_p$ values for single star and binary models (Table \ref{tab:single_models}. - \ref{tab:binary_models}.) within the same mass range, the single star progenitor is around 2-3 times more extended than the binary one. Moreover, as we fixed the additional nickel mass generated by the supernova explosion, Fig. \ref{fig:comp}. suggests that the initial nickel content of the binary progenitor models is much smaller than single-star ones with similar masses.

\begin{figure}[]
\centering
\includegraphics[width=0.48\textwidth]{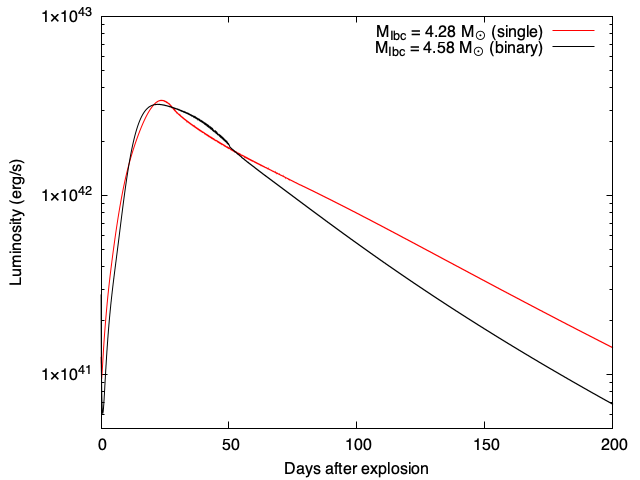}
\caption{Comparison of a single-star and a binary progenitor model calculated with $M_{Ibc}$ and similar peak luminosities. }
\label{fig:sim_lc}
\end{figure}

Note that we can generate similar peak luminosities for both single- and binary models if we explode binary progenitors with higher energies and nickel masses. For example, the $M_{Ibc} = 4.58 M_\odot$ binary model needed $3 \cdot 10^{51}$ erg and 0.25 $M_\odot$ to create a similar light curve maximum as the $M_{Ibc} = 4.28 M_\odot$ single-star model with our fix values of energy ($1.5 \cdot 10^{51}$) and nickel mass (0.1 $M_\odot$). Nevertheless, the global LC features show some differences (Fig. \ref{fig:sim_lc}.): e.g., the peak is broader for the binary models.

In further calculations, we use these non-interacting models as references and examine the effect of the interaction between the supernova ejecta and a thin, low-mass CSM shell. As all simulations show the same characteristic features, we demonstrate our results via a one-on-one particular model related to single- and binary progenitors with $4.28 M_\odot$ (Table \ref{tab:single_models}. S1) and $4.58 M_\odot$ (Table \ref{tab:binary_models}. B12) final mass, respectively.

\begin{figure}[]
\centering
\includegraphics[width=0.48\textwidth]{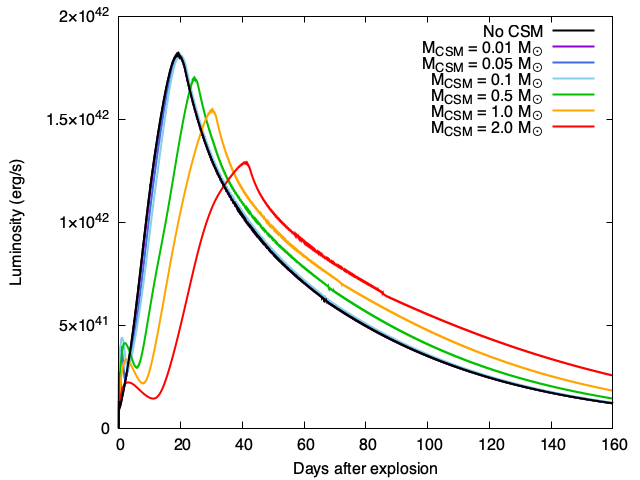}
\caption{The effect of different CSM mass on the bolometric light curves of single-star models. The black line represents the non-interacting reference model (Table \ref{tab:single_models}. S1). In contrast, the violet, dark blue, light blue, green, orange, and red illustrate the effect of circumstellar matter with a mass of 0.01, 0.05, 0.1, 0.5, 1, and 2 $M_\odot$, respectively.}
\label{fig:csm_mass}
\end{figure}

\begin{figure}[]
\centering
\includegraphics[width=0.48\textwidth]{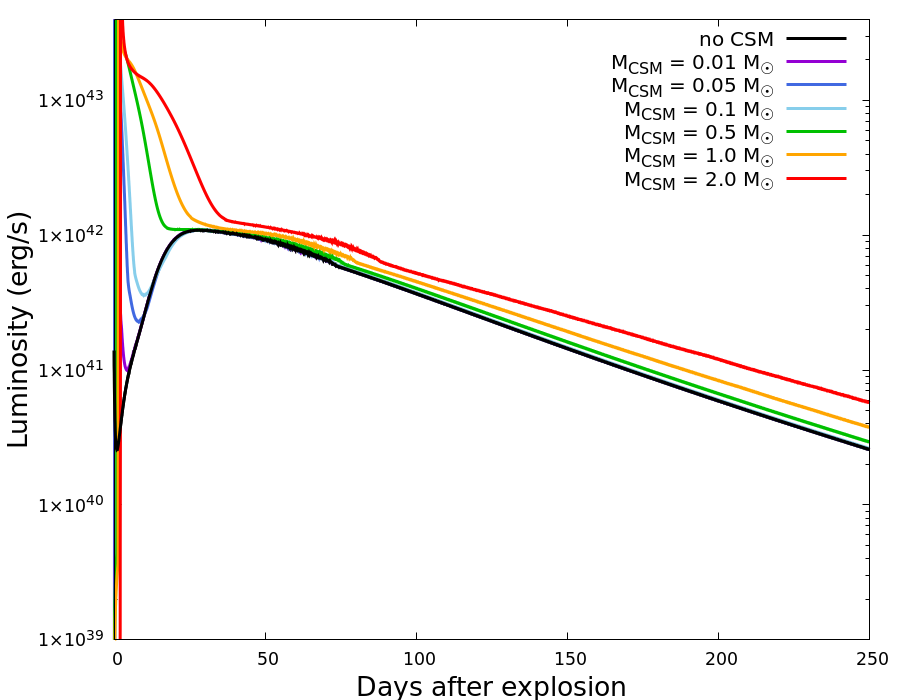}
\caption{The effect of different CSM mass on the bolometric light curves of binary models. The black line represents the non-interacting reference model (Table \ref{tab:binary_models}. B12). In contrast, the violet, dark blue, light blue, green, orange, and red illustrate the effect of circumstellar matter with a mass of 0.01, 0.05, 0.1, 0.5, 1, and 2 $M_\odot$, respectively.}
\label{fig:035_rad}
\end{figure}

First, we examine how the mass of the CSM affects the bolometric light curve while its radius is fixed as $10\cdot R_p$. Figure \ref{fig:csm_mass}. and \ref{fig:035_rad}. demonstrate our results for interacting single- and binary stars, respectively. Some general features can be seen in the generated LCs of both progenitor scenarios. Such as the appearance of a fast-declining early part similar to the first peak in Type IIb and IIP supernova light curves is also associated with a low-mass CSM envelope \citep{Chugai, Moriya11,Nagy1}. With increasing $M_{CSM}$, this early LC characteristic turns less luminous and broader, while the main light curve peak also changes significantly by the early bump. Moreover, with higher CSM mass, the late-time light curves more and more depart from the nickel-cobalt tail, and its slope becomes less steep.

On the other hand, the alteration of the second peak shows different attributes in the case of single-star and binary models. The interacting binary models preserve their rise time to the main maximum and remain about the same regardless of the interaction. Consequently, the early, fast-declining feature increasingly merges with the second LC peak as the CSM mass increases. These two components become inseparable at around $M_{CSM} = 0.5 M_{\odot}$. Furthermore, a plateau-like feature occurs at circa 80 days if we add at least 0.5 $M_{\odot}$ of circumstellar matter to our models. Note that the timescale and the luminosity of this feature would be more robust if a pure He-composed CSM had been taken into account (Figure \ref{fig:abund}.). By contrast, the single-star scenario shows two easily separable peaks: one generated by the CSM interaction, and the other originated from photon diffusion. Moreover, with increasing CSM masses, the rise time of the second peak shifts to later times and reduces its width and luminosity. Thus, the maximal luminosity of the two light curve features comes closer to each other. A possible explanation for this phenomenon could be related to the relative strength of radioactive decay and shock cooling emission. Namely, for low $M_{CSM}$, the main power source is the radioactive decay, while in a high-mass CSM medium, it is more likely to be the effect of slowly diffusing photons generated by the shock breakout \citep[see the case of SN 2023ixf,][]{Hiramatsu}.

\begin{figure}[]
\centering
\includegraphics[width=0.48\textwidth]{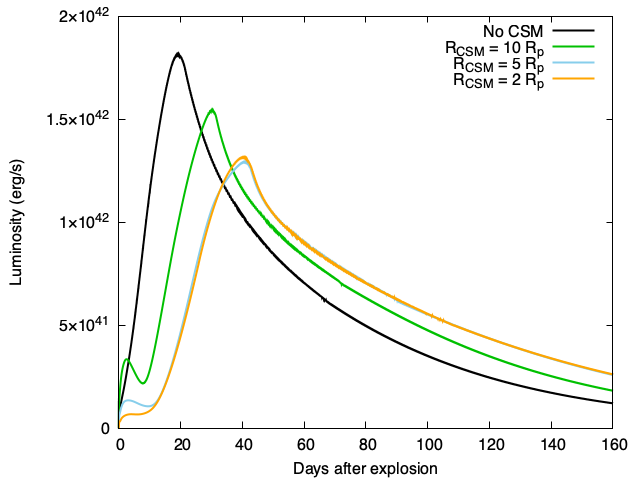}
\caption{The effect of different CSM radii on the bolometric light curves of single-star models. The black line represents the non-interacting reference models (Table \ref{tab:single_models} S1). In contrast, the green, light blue, and orange illustrate the effect of circumstellar matter with a radius of 10, 5, and 2 $R_p$, respectively.}
\label{fig:csm_rad}
\end{figure}

\begin{figure}[]
\centering
\includegraphics[width=0.48\textwidth]{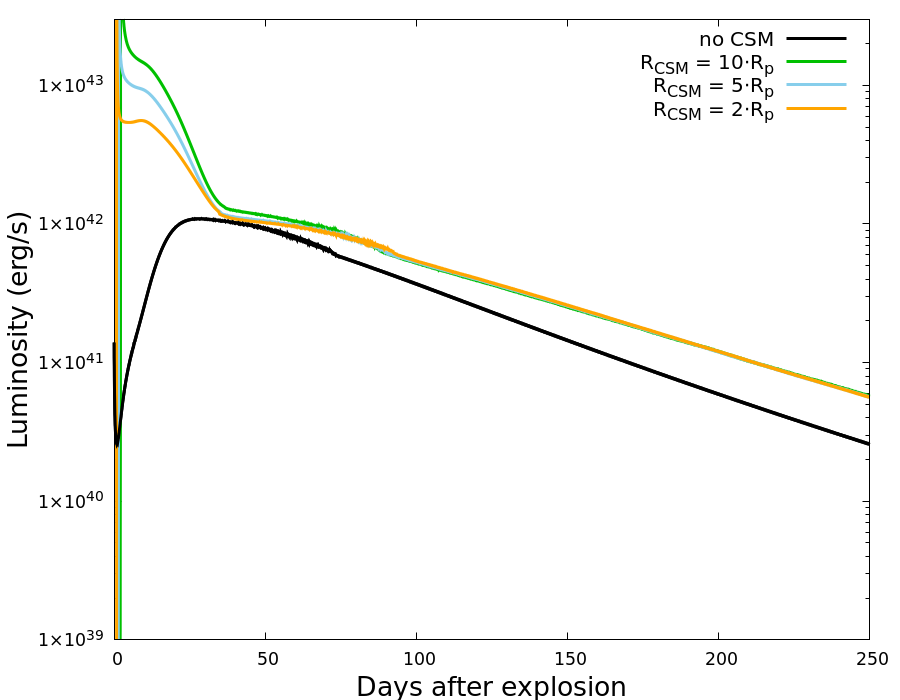}
\caption{The effect of different CSM radii on the bolometric light curves of binary models. The black line represents the non-interacting reference models (Table \ref{tab:binary_models}. B12). In contrast, the green, light blue, and orange illustrate the effect of circumstellar matter with a radius of 10, 5, and 2 $R_p$, respectively.}
\label{fig:035_mass}
\end{figure}

Besides CSM mass, the radius of the circumstellar matter may also affect the bolometric light curve. To test this scenario, we fixed $M_{CSM} = 2 M_\odot$ and only changed the value of $R_{CSM}$. As Fig. \ref{fig:csm_rad}. and \ref{fig:035_mass}. shows the effect of different CSM radii quite similar for both progenitor scenarios. In both cases, the width and the luminosity of the CSM interaction peak rise with increasing CSM radius. Meanwhile, the mean peak of the LC does not show any significant changes, except a slightly more luminous main peak for larger $R_{CSM}$ values.

To check the consistency of the interacting light curve models, we compare our results with the theoretical model published by \cite{Piro}. Here, we calculate $t_p$ and $L_p$ via Eq. 6 and 7. applying the different CSM model parameters, than compare with the simulation data. As a result we gained similar trends in both parameters suggesting that our models are consistent. Although, this qualitative check shows constancy, we are not able to determine  exact values for these parameters as \textbf{SNEC} define a non-constant opacity profile. Thus, we are only able to determine $t_p$ and $L_p$ within a factor of the average opacity. Nevertheless, if we estimate $\kappa$ for the different models as matching the theoretical calculations with model data, the average opacity varies between $0.03 - 0.2\ cm^2/g$ corresponding well with the integrated average opacity of stripped-envelope supernovae \citep{Nagy2}.

\section{Comparison with Observations}
To test our hypotheses that a low-mass, close CSM shell may cause strange light curve features (bumps, re-brightening, or steeper tail), we carefully select the reference explosion events. We chose three recent Type Ib/c supernovae (SN 2020oi, iPTF15dtg, and SN 2008D) classified as non-interacting SESNe and show some irregularities in their LCs. Moreover, we also advert that these selected transients should obtain similar bolometric light curve features as constructed by our CSM-interacting models. To compare synthesized light curves with observational data, we need to calculate the bolometric LCs of the picked supernovae. To do so, we use the same method published in \cite{Nagy}, which applies the trapezoidal rule for the integration over wavelength, assuming that the flux reaches zero at 2000 $\AA$. However, as \cite{Nakar} and \cite{Haynie} pointed out, the LC could reach its first maximum in higher frequencies than optical by incomplete thermalization around the shock breakout. Thus, the luminosity of the first peak could be underestimated: this is the case for SN 2008D, but also likely for the two other SNe \citep[see][for SN 2020oi]{Gagliano}.

Here, we mainly focused on the overall shape of the light curves as it may indicate whether the progenitor is a single star or it was part of a binary system. Thus, to determine the physical properties of these three supernovae was beyond the scope of this paper, we just tried to recreate their generic LC features and shift the observed and model light curves together. Thus, we used S1 (Table \ref{tab:single_models}.) single-star progenitor model for all three objects, while we adopt B2 and B12 (Table \ref{tab:binary_models}.) binary calculations for SN 2008D and SN 2020oi/iPTF15dtg, respectively. Nevertheless, we want to ensure that our shifting method does not cause any systematic error that can alter our conclusions. So, to get rid of the limitations of our model grid, we applied the Arnett model \citep{Arnett82}, where the relation between the ejecta mass and the explosion energy $(M/E_{exp}^2)$ approximately defines the width of the LC. Thus, if we scale the previously published explosion energies (SN 2008D: \cite{Tanaka}; iPTF15dtg: \cite{Taddia2}; SN 2020oi: \cite{Rho, Gagliano}) according to this relation, we only need to shift our model light curves up or down to compare with the supernova data.

One of our chosen supernovae: SN 2020oi was discovered on January 7. 2020. by the Zwicky Transient Facility \citep{Forster} in M100, which is a nearby galaxy with a distance of 16.22 Mpc \citep{Rho}. Although, this object was classified as a normal Type Ic, \cite{Horesh} and \cite{Maeda21} suggest a possible interaction between the supernova ejecta and a supposed dense circumstellar material around the progenitor. Moreover, \cite{Gagliano} even showed that SN 2020oi exploded in a binary system. Thus, this recent transient seemed to be perfect for testing our interacting binary models.

\begin{figure}
\centering
\includegraphics[width=.55\columnwidth]{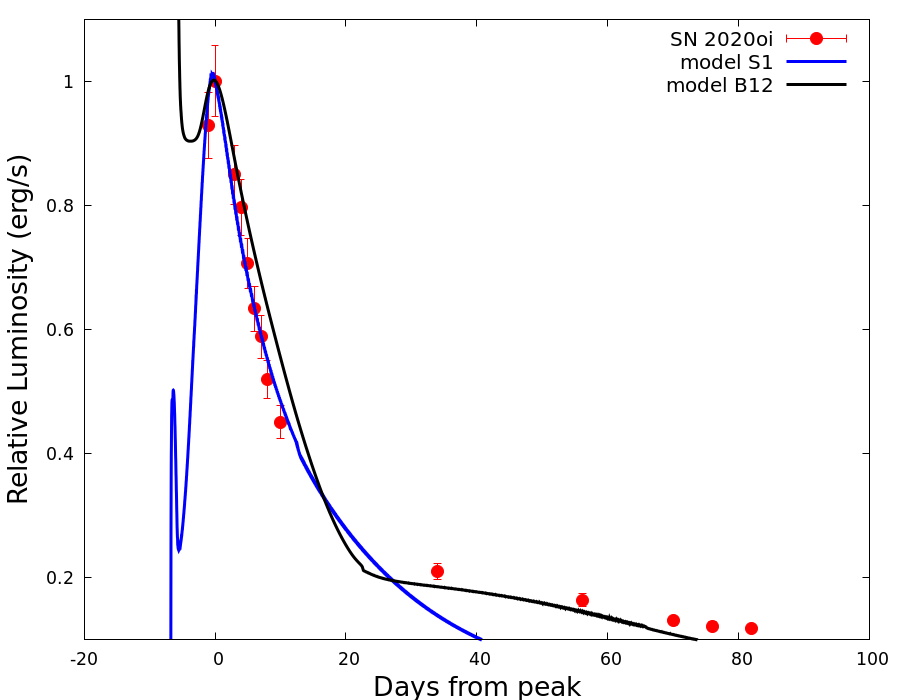}
\includegraphics[width=.55\columnwidth]{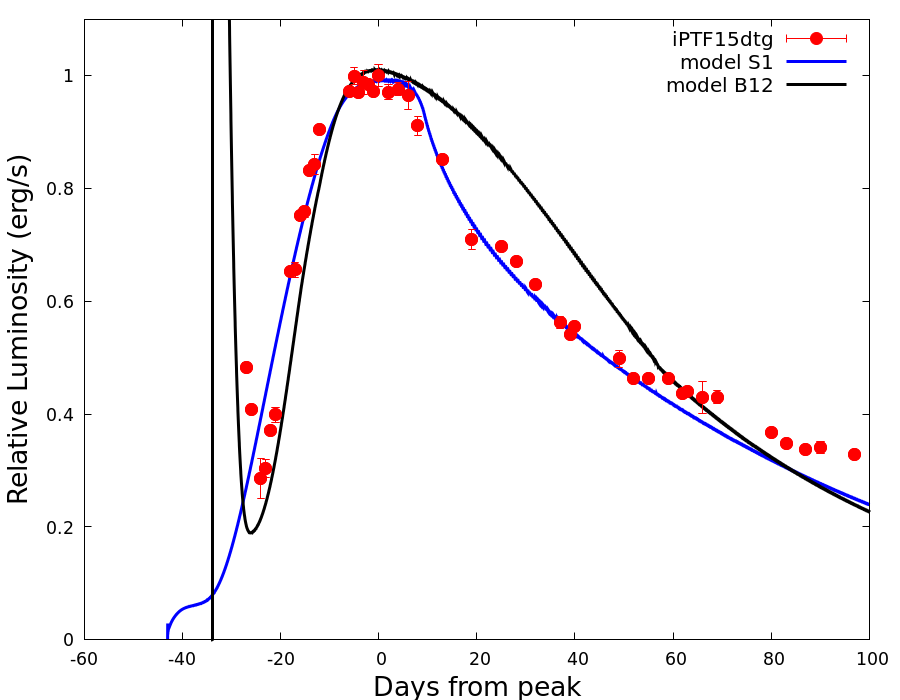}
\includegraphics[width=.55\columnwidth]{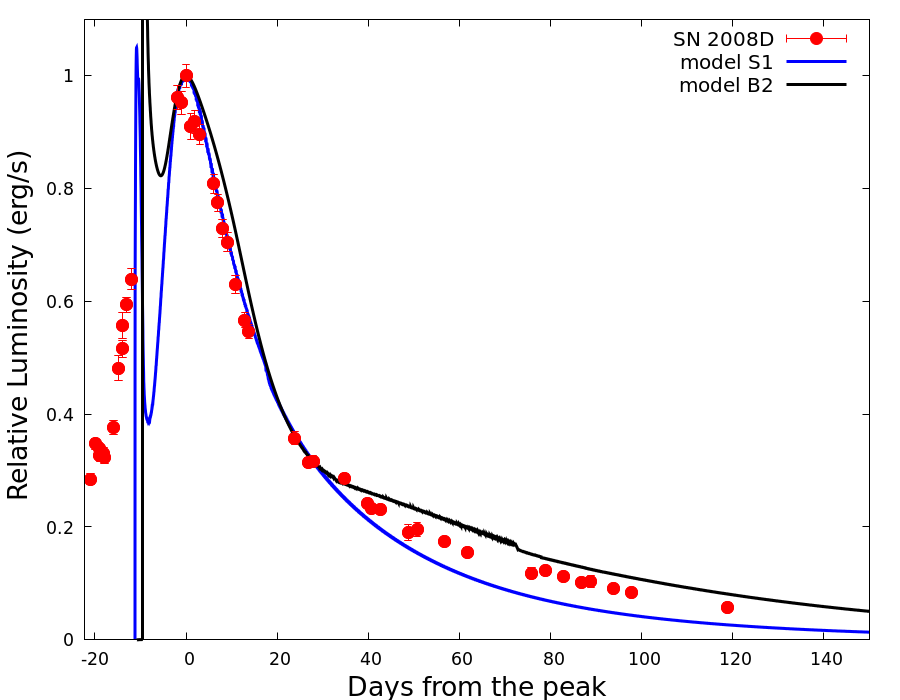}
\caption{Comparison of an interacting single-star (blue line) and an interacting binary (black line) model with the bolometric light curve (red circles) of SN 2020oi (top panel), iPTF15dtg (middle panel), and SN 200D (bottom panel), respectively.}
\label{fig:sn}
\end{figure}

In the top panel of Figure \ref{fig:sn}. we compare our interacting binary model related to $M_{CSM}= 2 M_{\odot}$ and $R_{CSM}=2\cdot R_p$  with the bolometric light curve of SN 2020oi generated from photometric data published by \citep{Rho}. To better see the resemblance between the two light curves, we raised the initial nickel mass of the hydrodynamic model to $0.2 M_{\odot}$. Again, we note that we did not try to recreate the explosion parameters in detail. We only scaled the luminosities together, to demonstrate how the shapes of the curves look relative to each other. As can be seen, the synthesized light curves follow the same temporal evolution as the bolometric LC of SN 2020oi. To make sure that our interacting single-star scenario is not capable of recreating the light variation of SN 2020oi properly, we try to find an acceptable model. We also show our most promising single-star model in the same panel of Fig. \ref{fig:sn}.  Considering our results and the findings of \cite{Rho, Horesh, Gagliano}, SN 2020oi could probably be an interacting SESN with a binary origin.

Our other chosen supernova was iPTF15dtg, discovered on November 7. 2015. by Palomar Transient Factory with the 96 Mpixel mosaic camera CFH12K \citep{Rahmer}. It was located in an anonymous galaxy with a luminosity distance of about 232.0 Mpc \citep{Taddia2}. Due to its high luminosity, iPTF15dtg was classified as a peculiar, slow-rising Type Ic supernova. However, \cite{Jin} suggests that this transient is an interacting supernova with an assumed $M_{CSM} = 0.05 - 0.15 M_\odot$ circumstellar matter that makes it an ideal test object for our single-star models.

In the middle panel of Figure \ref{fig:sn}. we compare our interacting single-star model related to  $M_{CSM}= 0.05 M_{\odot}$ and $R_{CSM}=5\cdot R_p$  with the bolometric light curve of the iPTF15dtg. Here, we also scaled the luminosities together, and we gained reasonably good agreement with the bolometric LC of iPTF15dtg. To rule out the binary origin, we create interacting binary models to find the most iPTF15dtg-like light curve shape also plotted to the corresponding figure. Considering our results, it seems more feasible that this supernova is originated from a single-star progenitor as our interacting binary models unable to create the rising part of the LC peak.

Finally, we used the Type Ib supernova, SN 2008D discovered on January 9. 2008 in NGC 2770 (d=27 Mpc) \citep{Soderberg}. It was originally detected with Swift X-ray Telescope as an X-ray transient and about 1.5 hours later appeared in the optical images. Later, theoretical studies recommend extremely different scenarios to explain the observational features of SN 2008D. Some of them suggest a compact, energetic, and aspherical single-star explosion, where the ejecta collides with the circumstellar matter \citep[e.g.][]{Scully}, while others prefer a binary origin \citep[e.g.][]{Brown}.

In the bottom panel of Figure \ref{fig:sn}. we compare our scaled interacting single-star model related to  $M_{CSM}= 0.1 M_{\odot}$ and $R_{CSM}=10\cdot R_p$ with the bolometric light curve of the SN 2008D. Albeit the light curve fails to follow the observed data, the shape of the modeled luminosities suggests that more detailed calculations may fit the light variation of SN 2008D properly. We also create interacting binary models to find if a binary-originated progenitor can produce a more closely SN 2008D-like LC (Fig. \ref{fig:sn}). The best fitting model is related to a $M_{Ibc}=3.23 M_{\odot}$ progenitor with a $M_{CSM}= 2 M_{\odot}$ and $R_{CSM}=2\cdot R_p$ circumstellar matter. The binary progenitor agrees with the light curve of SN 2008D after the maximum. However, since binary models can not form the early, slow-rising features of the LC peak, it seems unlikely that this explosion event originated from a binary progenitor. But, single-star models also have a limitation in this case. Without more detailed calculations, we only assert that SN 2008D shows single-progenitor-originated light curve features, which suggests an interacting single-star precursor by chance. Thus, further modeling will be necessary to evaluate this scenario and reveal the true nature of the progenitor and the circumstellar environment related to this supernova explosion.

\section{Conclusions}
We have investigated the effect of close, low-mass CSM shells around stripped-envelope supernova progenitors interacting with the SN ejecta by analyzing their bolometric LCs. Based on this systematic examination of the synthesized light curves, we found substantial evidence favoring our hypotheses. Namely, the strange light curve features of Type Ib/c supernovae could be explained by circumstellar interaction. Naturally, we are aware of the fact that without detailed spectral analysis, we can not be sure that our models generate typical Type Ib/c spectra. However, the low mass and low density of the CSM may suggest that the estimated configuration does not produce narrow He/C/O lines in the spectra and also could be an explanation for SESNe that changes their classification type in time.

We also find that the evolution of the bolometric light curves of our interacting single-star models is different than that of the binary progenitors, possibly due to their diverse initial $^{56}$Ni distribution. As presented above, the progenitor dependence can be mainly monitored at the early LCs: the main peak of the LC shifts and narrows for single stars, while the rise-time and width remain about the same for binary progenitors. On the other hand, this early behavior also could be a trace of an inflated progenitor, where the star has an extended, low-density outer envelope on top of a more compact and more massive inner region \citep[e.g.][]{Nagy1, Bersten12}. However, these inflated progenitor models do not significantly affect the nickel-cobalt tail, unlike some of our CSM models, since the former ones can not infer such high masses to modify the late-time LC features. Hence, a more massive circumstellar material is the key requirement to explain the observable luminosity excess (e.g., dumps or re-brightening) on the Ni-Co tail.

Nevertheless, our results suggest that the overall light curve features mainly depend on the compactness of the progenitor star regardless of the mass of its remaining He-layer, which shows good agreement with the finding of \cite{Woosley19}. Moreover, this phenomenon results in major differences between the LCs of the distinct progenitor scenarios, regardless of their similar maximum luminosity (Fig. \ref{fig:sim_lc}.). Thus, this may indicate that the pre-supernova evolution of the exploding star can be estimated from the general physical properties of Type Ib/c light curves.

Regarding the main objectives of this study, we compare our models with observational data. From an empirical point of view, the bolometric LC of stripped-envelope supernovae with bumps or re-brightening are quite heterogeneous. Some of them (e.g. SN 2008D, SN 2014L, iPTF15dtg) show a rapidly rising, narrow peak that fades as fast as it increases, while others (e.g. SN 2003dh, SN 2019dge, SN 2020oi) seem to miss the ascending part, but their LC peak is usually broader. Of course, the second group could be affected by an observational bias. However, as demonstrated earlier for SN 2020oi, it may suggest an interacting binary progenitor. On the other hand, the bolometric light curve features of the first group resemble our interacting single-star models despite the initial peak. However, it also could be an observational issue, as it seems to be a problem for some Type IIb supernovae, where we discover the supernova too late to catch the first peak that is otherwise assumed to be a common feature for these objects. Moreover, according to our simulations, the first peak can only be detected about a few days after the explosion, which leads to the non-detection of this feature in the case of certain Type Ib/c events. Thus, the absence of the first peak does not indicate the lack of CSM interaction.

Overall, our study indicates that a relatively thin, detached, dense circumstellar matter may explain the behavior of some Type Ib/c supernovae showing unusually light curve features. The gained models can recreate the basic LC characteristics of these events but are unable to determine the exact physical properties of the CSM via fitting observational data, which is mainly due to the simplified structure of our CSM model, as well as the complex nature of the hydrodynamical calculations.

\section*{Data Availability}
\textbf{MESA} and \textbf{SNEC}, the software used to produce the simulations for this paper, are fully open-source codes available at \url{https://github.com/MESAHub/} and \url{https://stellarcollapse.org/index.php/SNEC.html}, respectively. The majority of initial parameters to recreate our models are presented in tables. Data files, models, etc., will be shared with users upon reasonable request.

\begin{acknowledgements}
Special thanks to Takashi J. Moriya for the valuable suggestions and discussions that helped to improve this article.
This project is supported by NKFIH/OTKA PD-134434 and FK-134432 grants of the National Research, Development and Innovation (NRDI) Office of Hungary. This project has received funding from the HUN-REN Hungarian Research Network.
\end{acknowledgements}

\bibliographystyle{aa} 
\bibliography{aa.bib} 

\begin{thebibliography}{85}
\expandafter\ifx\csname natexlab\endcsname\relax\def\natexlab#1{#1}\fi

\bibitem[{{Andriesse} \& {Viotti}(1979)}]{Andriesse}
{Andriesse}, C.~D. \& {Viotti}, R. 1979, in Mass Loss and Evolution of O-Type Stars, ed. P.~S. {Conti} \& C.~W.~H. {De Loore}, Vol.~83, 47--50

\bibitem[{{Arnett}(1982)}]{Arnett82}
{Arnett}, W.~D. 1982, \apj, 253, 785

\bibitem[{Arnett \& Fu(1989)}]{Arnett}
Arnett, W.~D. \& Fu, A. 1989, ApJ, 340, 396

\bibitem[{{Benetti} {et~al.}(2018){Benetti}, {Zampieri}, {Pastorello}, {Cappellaro}, {Pumo}, {Elias-Rosa}, {Ochner}, {Terreran}, {Tomasella}, {Taubenberger}, {Turatto}, {Morales-Garoffolo}, {Harutyunyan}, \& {Tartaglia}}]{Benetti}
{Benetti}, S., {Zampieri}, L., {Pastorello}, A., {et~al.} 2018, \mnras, 476, 261

\bibitem[{{Bersten} {et~al.}(2011){Bersten}, {Benvenuto}, \& {Hamuy}}]{Bersten}
{Bersten}, M.~C., {Benvenuto}, O., \& {Hamuy}, M. 2011, \apj, 729, 61

\bibitem[{{Bersten} {et~al.}(2012){Bersten}, {Benvenuto}, {Nomoto}, {Ergon}, {Folatelli}, {Sollerman}, {Benetti}, {Botticella}, {Fraser}, {Kotak}, {Maeda}, {Ochner}, \& {Tomasella}}]{Bersten12}
{Bersten}, M.~C., {Benvenuto}, O.~G., {Nomoto}, K., {et~al.} 2012, \apj, 757, 31

\bibitem[{{Bersten} {et~al.}(2013){Bersten}, {Tanaka}, {Tominaga}, {Benvenuto}, \& {Nomoto}}]{Bersten1}
{Bersten}, M.~C., {Tanaka}, M., {Tominaga}, N., {Benvenuto}, O.~G., \& {Nomoto}, K. 2013, \apj, 767, 143

\bibitem[{{Brown} \& {Lee}(2008)}]{Brown}
{Brown}, G.~E. \& {Lee}, C.-H. 2008, arXiv e-prints, arXiv:0810.0912

\bibitem[{{Cao} {et~al.}(2013){Cao}, {Kasliwal}, {Arcavi}, {Horesh}, {Hancock}, {Valenti}, {Cenko}, {Kulkarni}, {Gal-Yam}, {Gorbikov}, {Ofek}, {Sand}, {Yaron}, {Graham}, {Silverman}, {Wheeler}, {Marion}, {Walker}, {Mazzali}, {Howell}, {Li}, {Kong}, {Bloom}, {Nugent}, {Surace}, {Masci}, {Carpenter}, {Degenaar}, \& {Gelino}}]{Cao}
{Cao}, Y., {Kasliwal}, M.~M., {Arcavi}, I., {et~al.} 2013, \apjl, 775, L7

\bibitem[{{Chevalier} \& {Fransson}(2003)}]{Chevalier1}
{Chevalier}, R.~A. \& {Fransson}, C. 2003, in Supernovae and Gamma-Ray Bursters, ed. K.~{Weiler}, Vol. 598, 171--194

\bibitem[{{Chevalier} \& {Fransson}(2006)}]{Chevalier}
{Chevalier}, R.~A. \& {Fransson}, C. 2006, ApJ, 651, 381

\bibitem[{{Chugai} {et~al.}(2007){Chugai}, {Chevalier}, \& {Utrobin}}]{Chugai}
{Chugai}, N.~N., {Chevalier}, R.~A., \& {Utrobin}, V.~P. 2007, \apj, 662, 1136

\bibitem[{Clocchiatti \& Wheeler(1997)}]{Clocchiatti}
Clocchiatti, A. \& Wheeler, J.~C. 1997, ApJ, 491, 375

\bibitem[{{de Jager} {et~al.}(1988){de Jager}, {Nieuwenhuijzen}, \& {van der Hucht}}]{deJager}
{de Jager}, C., {Nieuwenhuijzen}, H., \& {van der Hucht}, K.~A. 1988, \aaps, 72, 259

\bibitem[{{Dessart} {et~al.}(2022){Dessart}, {Hillier}, \& {Kuncarayakti}}]{Dessart22}
{Dessart}, L., {Hillier}, D.~J., \& {Kuncarayakti}, H. 2022, \aap, 658, A130

\bibitem[{{Forster} {et~al.}(2020){Forster}, {Pignata}, {Bauer}, {Arredondo}, {Cabrera-Vives}, {Carrasco-Davis}, {Estevez}, {Huijse}, {Reyes}, {Reyes}, {Sanchez-Saez}, {Valenzuela}, {Castillo}, {Ruz-Mieres}, {Rodriguez-Mancini}, {Bauer}, {Catelan}, {Eyheramendy}, \& {Graham}}]{Forster}
{Forster}, F., {Pignata}, G., {Bauer}, F.~E., {et~al.} 2020, Transient Name Server Discovery Report, 2020-67, 1

\bibitem[{{Fraser}(2020)}]{Fraser}
{Fraser}, M. 2020, Royal Society Open Science, 7, 200467

\bibitem[{{Gagliano} {et~al.}(2022){Gagliano}, {Izzo}, {Kilpatrick}, {Mockler}, {Jacobson-Gal{\'a}n}, {Terreran}, {Dimitriadis}, {Zenati}, {Auchettl}, {Drout}, {Narayan}, {Foley}, {Margutti}, {Rest}, {Jones}, {Aganze}, {Aleo}, {Burgasser}, {Coulter}, {Gerasimov}, {Gall}, {Hjorth}, {Hsu}, {Magnier}, {Mandel}, {Piro}, {Rojas-Bravo}, {Siebert}, {Stacey}, {Stroh}, {Swift}, {Taggart}, {Tinyanont}, \& {Tinyanont}}]{Gagliano}
{Gagliano}, A., {Izzo}, L., {Kilpatrick}, C.~D., {et~al.} 2022, \apj, 924, 55

\bibitem[{{Georgy} {et~al.}(2009){Georgy}, {Meynet}, {Walder}, {Folini}, \& {Maeder}}]{Georgy}
{Georgy}, C., {Meynet}, G., {Walder}, R., {Folini}, D., \& {Maeder}, A. 2009, \aap, 502, 611

\bibitem[{{Glebbeek} {et~al.}(2009){Glebbeek}, {Gaburov}, {de Mink}, {Pols}, \& {Portegies Zwart}}]{Glebbeek}
{Glebbeek}, E., {Gaburov}, E., {de Mink}, S.~E., {Pols}, O.~R., \& {Portegies Zwart}, S.~F. 2009, \aap, 497, 255

\bibitem[{{Hachinger} {et~al.}(2012){Hachinger}, {Mazzali}, {Taubenberger}, {Hillebrandt}, {Nomoto}, \& {Sauer}}]{Hachinger}
{Hachinger}, S., {Mazzali}, P.~A., {Taubenberger}, S., {et~al.} 2012, \mnras, 422, 70

\bibitem[{{Haynie} \& {Piro}(2021)}]{Haynie}
{Haynie}, A. \& {Piro}, A.~L. 2021, \apj, 910, 128

\bibitem[{{Hiramatsu} {et~al.}(2023){Hiramatsu}, {Tsuna}, {Berger}, {Itagaki}, {Goldberg}, {Gomez}, {Kishalay}, {Hosseinzadeh}, {Bostroem}, {Brown}, {Arcavi}, {Bieryla}, {Blanchard}, {Esquerdo}, {Farah}, {Howell}, {Matsumoto}, {McCully}, {Newsome}, {Gonzalez}, {Pellegrino}, {Rhee}, {Terreran}, {Vink{\'o}}, \& {Wheeler}}]{Hiramatsu}
{Hiramatsu}, D., {Tsuna}, D., {Berger}, E., {et~al.} 2023, \apjl, 955, L8

\bibitem[{{Ho} {et~al.}(2019){Ho}, {Goldstein}, {Schulze}, {Khatami}, {Perley}, {Ergon}, {Gal-Yam}, {Corsi}, {Andreoni}, {Barbarino}, {Bellm}, {Blagorodnova}, {Bright}, {Burns}, {Cenko}, {Cunningham}, {De}, {Dekany}, {Dugas}, {Fender}, {Fransson}, {Fremling}, {Goldstein}, {Graham}, {Hale}, {Horesh}, {Hung}, {Kasliwal}, {Kuin}, {Kulkarni}, {Kupfer}, {Lunnan}, {Masci}, {Ngeow}, {Nugent}, {Ofek}, {Patterson}, {Petitpas}, {Rusholme}, {Sai}, {Sfaradi}, {Shupe}, {Sollerman}, {Soumagnac}, {Tachibana}, {Taddia}, {Walters}, {Wang}, {Yao}, \& {Zhang}}]{Ho}
{Ho}, A. Y.~Q., {Goldstein}, D.~A., {Schulze}, S., {et~al.} 2019, \apj, 887, 169

\bibitem[{{Horesh} {et~al.}(2020){Horesh}, {Sfaradi}, {Ergon}, {Barbarino}, {Sollerman}, {Moldon}, {Dobie}, {Schulze}, {P{\'e}rez-Torres}, {Williams}, {Fremling}, {Gal-Yam}, {Kulkarni}, {O'Brien}, {Lundqvist}, {Murphy}, {Fender}, {Anand}, {Belicki}, {Bellm}, {Coughlin}, {De}, {Golkhou}, {Graham}, {Green}, {Hankins}, {Kasliwal}, {Kupfer}, {Laher}, {Masci}, {Miller}, {Neill}, {Ofek}, {Perrott}, {Porter}, {Reiley}, {Rigault}, {Rodriguez}, {Rusholme}, {Shupe}, \& {Titterington}}]{Horesh}
{Horesh}, A., {Sfaradi}, I., {Ergon}, M., {et~al.} 2020, \apj, 903, 132

\bibitem[{{Jermyn} {et~al.}(2023){Jermyn}, {Bauer}, {Schwab}, {Farmer}, {Ball}, {Bellinger}, {Dotter}, {Joyce}, {Marchant}, {Mombarg}, {Wolf}, {Sunny Wong}, {Cinquegrana}, {Farrell}, {Smolec}, {Thoul}, {Cantiello}, {Herwig}, {Toloza}, {Bildsten}, {Townsend}, \& {Timmes}}]{Jermyn}
{Jermyn}, A.~S., {Bauer}, E.~B., {Schwab}, J., {et~al.} 2023, \apjs, 265, 15

\bibitem[{{Jin} {et~al.}(2021){Jin}, {Yoon}, \& {Blinnikov}}]{Jin}
{Jin}, H., {Yoon}, S.-C., \& {Blinnikov}, S. 2021, \apj, 910, 68

\bibitem[{{Jin} {et~al.}(2023){Jin}, {Yoon}, \& {Blinnikov}}]{Jin1}
{Jin}, H., {Yoon}, S.-C., \& {Blinnikov}, S. 2023, \apj, 950, 44

\bibitem[{{Jung} {et~al.}(2022){Jung}, {Yoon}, \& {Kim}}]{Jung}
{Jung}, M.-K., {Yoon}, S.-C., \& {Kim}, H.-J. 2022, \apj, 925, 216

\bibitem[{{Kashi} \& {Michaelis}(2021)}]{Kashi}
{Kashi}, A. \& {Michaelis}, A. 2021, Galaxies, 10, 4

\bibitem[{{Kochanek}(2019)}]{Kochanek}
{Kochanek}, C.~S. 2019, \mnras, 483, 3762

\bibitem[{{Kuncarayakti} {et~al.}(2022){Kuncarayakti}, {Maeda}, {Dessart}, {Nagao}, {Fulton}, {Guti{\'e}rrez}, {Huber}, {Young}, {Kotak}, {Mattila}, {Anderson}, {Ferrari}, {Folatelli}, {Gao}, {Magnier}, {Smith}, \& {Srivastav}}]{Kuncarayakti}
{Kuncarayakti}, H., {Maeda}, K., {Dessart}, L., {et~al.} 2022, \apjl, 941, L32

\bibitem[{{Kuncarayakti} {et~al.}(2023){Kuncarayakti}, {Sollerman}, {Izzo}, {Maeda}, {Yang}, {Schulze}, {Angus}, {Aubert}, {Auchettl}, {Della Valle}, {Dessart}, {Hinds}, {Kankare}, {Kawabata}, {Lundqvist}, {Nakaoka}, {Perley}, {Raimundo}, {Strotjohann}, {Taguchi}, {Cai}, {Charalampopoulos}, {Fang}, {Fraser}, {Guti{\'e}rrez}, {Imazawa}, {Kangas}, {Kawabata}, {Kotak}, {Kravtsov}, {Matilainen}, {Mattila}, {Moran}, {Murata}, {Salmaso}, {Anderson}, {Ashall}, {Bellm}, {Benetti}, {Chambers}, {Chen}, {Coughlin}, {De Colle}, {Fremling}, {Galbany}, {Gal-Yam}, {Gromadzki}, {Groom}, {Hajela}, {Inserra}, {Kasliwal}, {Mahabal}, {Martin-Carrillo}, {Moore}, {M{\"u}ller-Bravo}, {Nicholl}, {Ragosta}, {Riddle}, {Sharma}, {Srivastav}, {Stritzinger}, {Wold}, \& {Young}}]{Kuncarayakti1}
{Kuncarayakti}, H., {Sollerman}, J., {Izzo}, L., {et~al.} 2023, \aap, 678, A209

\bibitem[{{Kuriyama} \& {Shigeyama}(2020)}]{Kuriyama}
{Kuriyama}, N. \& {Shigeyama}, T. 2020, \aap, 635, A127

\bibitem[{{Limongi}(2017)}]{Limongi}
{Limongi}, M. 2017, in Handbook of Supernovae, ed. A.~W. {Alsabti} \& P.~{Murdin}, 513

\bibitem[{Lyman {et~al.}(2016)Lyman, Bersier, James, Mazzali, Eldridge, Fraser, \& Pian}]{Lyman}
Lyman, J.~D., Bersier, D., James, P.~A., {et~al.} 2016, MNRAS, 457, 328–350

\bibitem[{{Maeda} {et~al.}(2021){Maeda}, {Chandra}, {Matsuoka}, {Ryder}, {Moriya}, {Kuncarayakti}, {Lee}, {Kundu}, {Patnaude}, {Saito}, \& {Folatelli}}]{Maeda21}
{Maeda}, K., {Chandra}, P., {Matsuoka}, T., {et~al.} 2021, \apj, 918, 34

\bibitem[{{Maeda} \& {Moriya}(2022)}]{Maeda22}
{Maeda}, K. \& {Moriya}, T.~J. 2022, \apj, 927, 25

\bibitem[{{Mazzali} {et~al.}(2016){Mazzali}, {Sullivan}, {Pian}, {Greiner}, \& {Kann}}]{Mazzali}
{Mazzali}, P.~A., {Sullivan}, M., {Pian}, E., {Greiner}, J., \& {Kann}, D.~A. 2016, \mnras, 458, 3455

\bibitem[{{Mcley} \& {Soker}(2014)}]{Mcley}
{Mcley}, L. \& {Soker}, N. 2014, \mnras, 445, 2492

\bibitem[{{Moriya} {et~al.}(2011){Moriya}, {Tominaga}, {Blinnikov}, {Baklanov}, \& {Sorokina}}]{Moriya11}
{Moriya}, T., {Tominaga}, N., {Blinnikov}, S.~I., {Baklanov}, P.~V., \& {Sorokina}, E.~I. 2011, \mnras, 415, 199

\bibitem[{{Moriya} \& {Maeda}(2012)}]{Moriya2}
{Moriya}, T.~J. \& {Maeda}, K. 2012, \apjl, 756, L22

\bibitem[{{Morozova} {et~al.}(2015){Morozova}, {Piro}, {Renzo}, {Ott}, {Clausen}, {Couch}, {Ellis}, \& {Roberts}}]{Morozova}
{Morozova}, V., {Piro}, A.~L., {Renzo}, M., {et~al.} 2015, \apj, 814, 63

\bibitem[{{M{\"u}ller} {et~al.}(2017){M{\"u}ller}, {Prieto}, {Pejcha}, \& {Clocchiatti}}]{Muller}
{M{\"u}ller}, T., {Prieto}, J.~L., {Pejcha}, O., \& {Clocchiatti}, A. 2017, \apj, 841, 127

\bibitem[{Nagy(2018)}]{Nagy2}
Nagy, A.~P. 2018, ApJ, 862, 143

\bibitem[{Nagy {et~al.}(2014)Nagy, Ordasi, Vinkó, \& Wheeler}]{Nagy}
Nagy, A.~P., Ordasi, A., Vinkó, J., \& Wheeler, J.~C. 2014, A\&A, 571, A77

\bibitem[{Nagy \& Vinkó(2016)}]{Nagy1}
Nagy, A.~P. \& Vinkó, J. 2016, A\&A, 589, A53

\bibitem[{{Nakar} \& {Sari}(2010)}]{Nakar}
{Nakar}, E. \& {Sari}, R. 2010, \apj, 725, 904

\bibitem[{{Nugis} \& {Lamers}(2000)}]{Nugis}
{Nugis}, T. \& {Lamers}, H.~J.~G.~L.~M. 2000, \aap, 360, 227

\bibitem[{{Ouchi} {et~al.}(2021){Ouchi}, {Maeda}, {Anderson}, \& {Sawada}}]{Ouchi}
{Ouchi}, R., {Maeda}, K., {Anderson}, J.~P., \& {Sawada}, R. 2021, \apj, 922, 141

\bibitem[{{Paczynski}(1983)}]{Paczynski}
{Paczynski}, B. 1983, \apj, 267, 315

\bibitem[{{Pastorello} {et~al.}(2007){Pastorello}, {Smartt}, {Mattila}, {Eldridge}, {Young}, {Itagaki}, {Yamaoka}, {Navasardyan}, {Valenti}, {Patat}, {Agnoletto}, {Augusteijn}, {Benetti}, {Cappellaro}, {Boles}, {Bonnet-Bidaud}, {Botticella}, {Bufano}, {Cao}, {Deng}, {Dennefeld}, {Elias-Rosa}, {Harutyunyan}, {Keenan}, {Iijima}, {Lorenzi}, {Mazzali}, {Meng}, {Nakano}, {Nielsen}, {Smoker}, {Stanishev}, {Turatto}, {Xu}, \& {Zampieri}}]{Pastorello}
{Pastorello}, A., {Smartt}, S.~J., {Mattila}, S., {et~al.} 2007, \nat, 447, 829

\bibitem[{{Pauli} {et~al.}(2023){Pauli}, {Oskinova}, {Hamann}, {Bowman}, {Todt}, {Shenar}, {Sander}, {Erba}, {G{\'o}mez-Gonz{\'a}lez}, {Kehrig}, {Klencki}, {Kuiper}, {Mehner}, {de Mink}, {Oey}, {Ramachandran}, {Schootemeijer}, {Reyero Serantes}, \& {Wofford}}]{Pauli}
{Pauli}, D., {Oskinova}, L.~M., {Hamann}, W.~R., {et~al.} 2023, \aap, 673, A40

\bibitem[{{Paxton} {et~al.}(2011){Paxton}, {Bildsten}, {Dotter}, {Herwig}, {Lesaffre}, \& {Timmes}}]{Paxton1}
{Paxton}, B., {Bildsten}, L., {Dotter}, A., {et~al.} 2011, \apjs, 192, 3

\bibitem[{{Paxton} {et~al.}(2013){Paxton}, {Cantiello}, {Arras}, {Bildsten}, {Brown}, {Dotter}, {Mankovich}, {Montgomery}, {Stello}, {Timmes}, \& {Townsend}}]{Paxton2}
{Paxton}, B., {Cantiello}, M., {Arras}, P., {et~al.} 2013, \apjs, 208, 4

\bibitem[{{Paxton} {et~al.}(2015){Paxton}, {Marchant}, {Schwab}, {Bauer}, {Bildsten}, {Cantiello}, {Dessart}, {Farmer}, {Hu}, {Langer}, {Townsend}, {Townsley}, \& {Timmes}}]{Paxton3}
{Paxton}, B., {Marchant}, P., {Schwab}, J., {et~al.} 2015, \apjs, 220, 15

\bibitem[{{Paxton} {et~al.}(2018){Paxton}, {Schwab}, {Bauer}, {Bildsten}, {Blinnikov}, {Duffell}, {Farmer}, {Goldberg}, {Marchant}, {Sorokina}, {Thoul}, {Townsend}, \& {Timmes}}]{Paxton4}
{Paxton}, B., {Schwab}, J., {Bauer}, E.~B., {et~al.} 2018, \apjs, 234, 34

\bibitem[{{Paxton} {et~al.}(2019){Paxton}, {Smolec}, {Schwab}, {Gautschy}, {Bildsten}, {Cantiello}, {Dotter}, {Farmer}, {Goldberg}, {Jermyn}, {Kanbur}, {Marchant}, {Thoul}, {Townsend}, {Wolf}, {Zhang}, \& {Timmes}}]{Paxton5}
{Paxton}, B., {Smolec}, R., {Schwab}, J., {et~al.} 2019, \apjs, 243, 10

\bibitem[{{Petrovic} {et~al.}(2005){Petrovic}, {Langer}, \& {van der Hucht}}]{Petrovic}
{Petrovic}, J., {Langer}, N., \& {van der Hucht}, K.~A. 2005, \aap, 435, 1013

\bibitem[{{Piro}(2015)}]{Piro}
{Piro}, A.~L. 2015, \apjl, 808, L51

\bibitem[{{Quataert} {et~al.}(2016){Quataert}, {Fern{\'a}ndez}, {Kasen}, {Klion}, \& {Paxton}}]{Quataert}
{Quataert}, E., {Fern{\'a}ndez}, R., {Kasen}, D., {Klion}, H., \& {Paxton}, B. 2016, \mnras, 458, 1214

\bibitem[{{Rahmer} {et~al.}(2008){Rahmer}, {Smith}, {Velur}, {Hale}, {Law}, {Bui}, {Petrie}, \& {Dekany}}]{Rahmer}
{Rahmer}, G., {Smith}, R., {Velur}, V., {et~al.} 2008, in Society of Photo-Optical Instrumentation Engineers (SPIE) Conference Series, Vol. 7014, Ground-based and Airborne Instrumentation for Astronomy II, ed. I.~S. {McLean} \& M.~M. {Casali}, 70144Y

\bibitem[{{Reimers}(1975)}]{Reimers}
{Reimers}, D. 1975, in Problems in stellar atmospheres and envelopes., 229--256

\bibitem[{{Rho} {et~al.}(2021){Rho}, {Evans}, {Geballe}, {Banerjee}, {Hoeflich}, {Shahbandeh}, {Valenti}, {Yoon}, {Jin}, {Williamson}, {Modjaz}, {Hiramatsu}, {Howell}, {Pellegrino}, {Vink{\'o}}, {Cartier}, {Burke}, {McCully}, {An}, {Cha}, {Pritchard}, {Wang}, {Andrews}, {Galbany}, {Van Dyk}, {Graham}, {Blinnikov}, {Joshi}, {P{\'a}l}, {Kriskovics}, {Ordasi}, {Szakats}, {Vida}, {Chen}, {Li}, {Zhang}, \& {Yan}}]{Rho}
{Rho}, J., {Evans}, A., {Geballe}, T.~R., {et~al.} 2021, \apj, 908, 232

\bibitem[{{Rodr{\'\i}guez} {et~al.}(2023){Rodr{\'\i}guez}, {Maoz}, \& {Nakar}}]{Rodriguez}
{Rodr{\'\i}guez}, {\'O}., {Maoz}, D., \& {Nakar}, E. 2023, \apj, 955, 71

\bibitem[{{Sana} {et~al.}(2012){Sana}, {de Mink}, {de Koter}, {Langer}, {Evans}, {Gieles}, {Gosset}, {Izzard}, {Le Bouquin}, \& {Schneider}}]{Sana}
{Sana}, H., {de Mink}, S.~E., {de Koter}, A., {et~al.} 2012, Science, 337, 444

\bibitem[{{Scully} {et~al.}(2023){Scully}, {Matzner}, \& {Yalinewich}}]{Scully}
{Scully}, B., {Matzner}, C.~D., \& {Yalinewich}, A. 2023, \mnras, 525, 1562

\bibitem[{{Shao} \& {Li}(2016)}]{Shao}
{Shao}, Y. \& {Li}, X.-D. 2016, \apj, 833, 108

\bibitem[{{Soderberg} {et~al.}(2008){Soderberg}, {Berger}, {Page}, {Schady}, {Parrent}, {Pooley}, {Wang}, {Ofek}, {Cucchiara}, {Rau}, {Waxman}, {Simon}, {Bock}, {Milne}, {Page}, {Barentine}, {Barthelmy}, {Beardmore}, {Bietenholz}, {Brown}, {Burrows}, {Burrows}, {Byrngelson}, {Cenko}, {Chandra}, {Cummings}, {Fox}, {Gal-Yam}, {Gehrels}, {Immler}, {Kasliwal}, {Kong}, {Krimm}, {Kulkarni}, {Maccarone}, {M{\'e}sz{\'a}ros}, {Nakar}, {O'Brien}, {Overzier}, {de Pasquale}, {Racusin}, {Rea}, \& {York}}]{Soderberg}
{Soderberg}, A.~M., {Berger}, E., {Page}, K.~L., {et~al.} 2008, \nat, 453, 469

\bibitem[{Sollerman {et~al.}(2020)Sollerman, C., Barbarino, Fremling, Horesh, Kool, Schulze, Sfaradi, Yang, Bellm, Burruss, Cunningham, De, Drake, Golkhou, Green, Kasliwal, Kulkarni, Kupfer, Laher, Masci, Rodriguez, Rusholme, Williams, Yan, \& Zolkower}]{Sollerman}
Sollerman, J., C., F., Barbarino, C., {et~al.} 2020, A\&A, 643, A79

\bibitem[{{Soraisam} {et~al.}(2022){Soraisam}, {Matheson}, {Lee}, {Saha}, {Narayan}, {Wolf}, {Scott}, {Figuereo}, {Nu{\~n}ez}, {McKinnon}, {Guhathakurta}, {Brink}, {Filippenko}, \& {Smith}}]{Soraisam}
{Soraisam}, M., {Matheson}, T., {Lee}, C.-H., {et~al.} 2022, \apjl, 926, L11

\bibitem[{{Strotjohann} {et~al.}(2021){Strotjohann}, {Ofek}, {Gal-Yam}, {Bruch}, {Schulze}, {Shaviv}, {Sollerman}, {Filippenko}, {Yaron}, {Fremling}, {Nordin}, {Kool}, {Perley}, {Ho}, {Yang}, {Yao}, {Soumagnac}, {Graham}, {Barbarino}, {Tartaglia}, {De}, {Goldstein}, {Cook}, {Brink}, {Taggart}, {Yan}, {Lunnan}, {Kasliwal}, {Kulkarni}, {Nugent}, {Masci}, {Rosnet}, {Adams}, {Andreoni}, {Bagdasaryan}, {Bellm}, {Burdge}, {Duev}, {Dugas}, {Frederick}, {Goldwasser}, {Hankins}, {Irani}, {Karambelkar}, {Kupfer}, {Liang}, {Neill}, {Porter}, {Riddle}, {Sharma}, {Short}, {Taddia}, {Tzanidakis}, {van Roestel}, {Walters}, \& {Zhuang}}]{Strotjohann}
{Strotjohann}, N.~L., {Ofek}, E.~O., {Gal-Yam}, A., {et~al.} 2021, \apj, 907, 99

\bibitem[{{Taddia} {et~al.}(2016){Taddia}, {Fremling}, {Sollerman}, {Corsi}, {Gal-Yam}, {Karamehmetoglu}, {Lunnan}, {Bue}, {Ergon}, {Kasliwal}, {Vreeswijk}, \& {Wozniak}}]{Taddia2}
{Taddia}, F., {Fremling}, C., {Sollerman}, J., {et~al.} 2016, \aap, 592, A89

\bibitem[{Taddia {et~al.}(2015)Taddia, Sollerman, Leloudas, Stritzinger, Valenti, Galbany, Kessler, Schneider, \& Wheeler}]{Taddia}
Taddia, F., Sollerman, J., Leloudas, G., {et~al.} 2015, A\&A, 547, A60

\bibitem[{{Taddia} {et~al.}(2018){Taddia}, {Stritzinger}, {Bersten}, {Baron}, {Burns}, {Contreras}, {Holmbo}, {Hsiao}, {Morrell}, {Phillips}, {Sollerman}, \& {Suntzeff}}]{Taddia1}
{Taddia}, F., {Stritzinger}, M.~D., {Bersten}, M., {et~al.} 2018, \aap, 609, A136

\bibitem[{{Takei} {et~al.}(2024){Takei}, {Tsuna}, {Ko}, \& {Shigeyama}}]{Takei}
{Takei}, Y., {Tsuna}, D., {Ko}, T., \& {Shigeyama}, T. 2024, \apj, 961, 67

\bibitem[{{Tanaka} {et~al.}(2009){Tanaka}, {Tominaga}, {Nomoto}, {Valenti}, {Sahu}, {Minezaki}, {Yoshii}, {Yoshida}, {Anupama}, {Benetti}, {Chincarini}, {Della Valle}, {Mazzali}, \& {Pian}}]{Tanaka}
{Tanaka}, M., {Tominaga}, N., {Nomoto}, K., {et~al.} 2009, \apj, 692, 1131

\bibitem[{{Tsuna} \& {Takei}(2023)}]{Tsuna}
{Tsuna}, D. \& {Takei}, Y. 2023, \pasj, 75, L19

\bibitem[{{Vamvatira-Nakou} {et~al.}(2016){Vamvatira-Nakou}, {Hutsem{\'e}kers}, {Royer}, {Waelkens}, {Groenewegen}, \& {Barlow}}]{Vamcatira}
{Vamvatira-Nakou}, C., {Hutsem{\'e}kers}, D., {Royer}, P., {et~al.} 2016, \aap, 588, A92

\bibitem[{{Vink}(2022)}]{Vink}
{Vink}, J.~S. 2022, \araa, 60, 203

\bibitem[{{Vink} {et~al.}(2001){Vink}, {de Koter}, \& {Lamers}}]{Vink1}
{Vink}, J.~S., {de Koter}, A., \& {Lamers}, H.~J.~G.~L.~M. 2001, \aap, 369, 574

\bibitem[{{Wheeler} {et~al.}(2017){Wheeler}, {Chatzopoulos}, {Vink{\'o}}, \& {Tuminello}}]{Wheeler2}
{Wheeler}, J.~C., {Chatzopoulos}, E., {Vink{\'o}}, J., \& {Tuminello}, R. 2017, \apjl, 851, L14

\bibitem[{{Woosley}(2019)}]{Woosley19}
{Woosley}, S.~E. 2019, \apj, 878, 49

\bibitem[{{Woosley} {et~al.}(2004){Woosley}, {Heger}, {Cumming}, {Hoffman}, {Pruet}, {Rauscher}, {Fisker}, {Schatz}, {Brown}, \& {Wiescher}}]{Woosley1}
{Woosley}, S.~E., {Heger}, A., {Cumming}, A., {et~al.} 2004, \apjs, 151, 75

\bibitem[{{Woosley} {et~al.}(2021){Woosley}, {Sukhbold}, \& {Kasen}}]{Woosley}
{Woosley}, S.~E., {Sukhbold}, T., \& {Kasen}, D.~N. 2021, \apj, 913, 145

\end{thebibliography}

\end{document}